\documentclass[fleqn,11pt,twoside]{article}

\usepackage[english]{babel}
\usepackage[utf8]{inputenc}
\usepackage[T1]{fontenc}

\usepackage{amsmath}
\usepackage{amssymb}
\usepackage{mathtools}
\usepackage{amsthm}
\usepackage{xcolor}
\usepackage{epsfig}
\usepackage{subfigure}
\usepackage{graphicx}
\usepackage{latexsym}
\usepackage{lscape}
\usepackage{physics}
\usepackage{braket}
\usepackage{cleveref}
\usepackage{tikz}
\usetikzlibrary{calc}

\newcommand{\Pj}{\mathbb{CP}}

\newcommand{\R}{\mathbb{R}}
\newcommand{\Z}{\mathbb{Z}}
\newcommand{\N}{\mathbb{N}}

\newcommand{\Cp}{\mathbb{C}}

\renewcommand{\vec}[1]{\mathbf{#1}}

\renewcommand{\epsilon}{\varepsilon}
\renewcommand{\imath}{\mathrm{i}}

\allowdisplaybreaks
\makeatletter 
\renewcommand{\pdv}[2]{\begingroup 
\@tempswafalse\toks@={}\count@=\z@ 
\@for\next:=#2\do 
{\expandafter\check@var\next\@nil
 \advance\count@\der@exp 
 \if@tempswa 
   \toks@=\expandafter{\the\toks@\,}% 
 \else 
   \@tempswatrue 
 \fi 
 \toks@=\expandafter{\the\expandafter\toks@\expandafter\partial\der@var}}% 
\frac{\partial\ifnum\count@=\@ne\else^{\number\count@}\fi#1}{\the\toks@}% 
\endgroup} 
\def\check@var{\@ifstar{\mult@var}{\one@var}} 
\def\mult@var#1#2\@nil{\def\der@var{#2^{#1}}\def\der@exp{#1}} 
\def\one@var#1\@nil{\def\der@var{#1}\chardef\der@exp\@ne} 

\makeatletter
\newcommand{\copyrightnote}[2]{{\renewcommand{\thefootnote}{}
 \footnotetext{\small\it
\begin{flushleft}
 \copyright \ #1   #2  
\end{flushleft}}}}

\newcommand{\Name}[1]{\begin{flushleft}
                       \LARGE \bf #1
                       \end{flushleft}\vspace{-3mm}}

\newcommand{\Author}[1]{\begin{flushleft}
                       \it #1 \end{flushleft}}

\newcommand{\Address}[1]{\begin{flushleft}
                       \it #1 \end{flushleft}}

\newcommand{\Date}[1]{\begin{flushleft}
                      \small  \it #1 \end{flushleft}}

\newcommand{\evenhead}{Author \ name}
\newcommand{\oddhead}{Article \ name}

\renewcommand{\@evenhead}{
\hspace*{-3pt}\raisebox{-15pt}[\headheight][0pt]{\vbox{\hbox to \textwidth
{\thepage \hfil \evenhead}\vskip4pt \hrule}}}
\renewcommand{\@oddhead}{
\hspace*{-3pt}\raisebox{-15pt}[\headheight][0pt]{\vbox{\hbox to \textwidth
{\oddhead \hfil \thepage}\vskip4pt\hrule}}}
\renewcommand{\@evenfoot}{}
\renewcommand{\@oddfoot}{}

\long\def\@makecaption#1#2{%
  \vskip\abovecaptionskip
  \sbox\@tempboxa{\small \textbf{#1.}\ \ #2}%
  \ifdim \wd\@tempboxa >\hsize
    {\small \textbf{#1.}\ \ #2}\par
  \else
    \global \@minipagefalse
    \hb@xt@\hsize{\hfil\box\@tempboxa\hfil}%
  \fi
  \vskip\belowcaptionskip}

\newcommand{\JNMPnumberwithin}[3][\arabic]{%
  \@ifundefined{c@#2}{\@nocounterr{#2}}{%
    \@ifundefined{c@#3}{\@nocnterr{#3}}{%
      \@addtoreset{#2}{#3}%
      \@xp\xdef\csname the#2\endcsname{%
        \@xp\@nx\csname the#3\endcsname .\@nx#1{#2}}}}%
}

\newcommand{\resetfootnoterule} {
  \renewcommand\footnoterule{%
  \kern-3\p@
  \hrule\@width.4\columnwidth
  \kern2.6\p@}
}

\renewcommand{\footnoterule}{}

\makeatother

\theoremstyle{plain}
\newtheorem{theorem}{Theorem}[section]

\newtheorem{corollary}[theorem]{Corollary}

\theoremstyle{definition}
\newtheorem{definition}[theorem]{Definition}
\newtheorem{notation}[theorem]{Notation}

\theoremstyle{remark}
\newtheorem{remark}[theorem]{Remark}
\theoremstyle{remark}

\newtheorem*{conjecture*}{Conjecture}

\usepackage[backend=biber,sorting=nyt,maxnames=99,giveninits=true,isbn=false,doi=false,url=false,citestyle=numeric-comp]{biblatex}
\begin{document}

\renewcommand{\evenhead}{ {\LARGE\textcolor{blue!10!black!40!green}{{\sf \ \ \ ]ocnmp[}}}\strut\hfill G Gubbiotti}
\renewcommand{\oddhead}{ {\LARGE\textcolor{blue!10!black!40!green}{{\sf ]ocnmp[}}}\ \ \ \ \  Algebraic entropy for systems of quad equations}

%%%% Matter for the first page 
\thispagestyle{empty}
\newcommand{\FistPageHead}[3]{
\begin{flushleft}
\raisebox{8mm}[0pt][0pt]
{\footnotesize \sf
\parbox{150mm}{{Open Communications in Nonlinear Mathematical Physics}\ \ \ \ {\LARGE\textcolor{blue!10!black!40!green}{]ocnmp[}}
\quad Special Issue 1, 2024\ \  pp
#2\hfill {\sc #3}}}\vspace{-13mm}
\end{flushleft}}

\FistPageHead{1}{\pageref{firstpage}--\pageref{lastpage}}{ \ \ }

\strut\hfill

\strut\hfill

\copyrightnote{The author(s). Distributed under a Creative Commons Attribution 4.0 International License}

\begin{center}
%{\Large  {\sf This article is part of a Special Issue in Memory of Professor Decio Levi}}
{  {\bf This article is part of an OCNMP Special Issue\\ 
\smallskip
in Memory of Professor Decio Levi}}
\end{center}

\smallskip

\Name{Algebraic Entropy for Systems of Quad Equations}

\Author{Giorgio Gubbiotti}

\Address{Dipartimento di Matematica ``Federigo Enriques'', Universit\`a degli
Studi di Milano, Via C. Saldini 50, 20133 Milano, Italy \& INFN Sezione Milano,
Via G. Celoria 16, 20133 Milano, Italy, email: giorgio.gubbiotti\@unimi.it}

\Date{Received July 25, 2023; Accepted February 5, 2024}

\setcounter{equation}{0}

\begin{abstract}
    \noindent
    In this work I discuss briefly the calculation
    of the algebraic entropy for systems of quad equations.
    In particular, I observe that since systems of multilinear
    equations can have algebraic solution, in some cases one
    might need to restrict the direction of evolution only
    to the pair of vertices yielding a birational evolution.
    Some examples from the existing literature are presented
    and discussed within this framework.
\end{abstract}

\label{firstpage}

%%%% The Article text starts here

\section{Introduction}

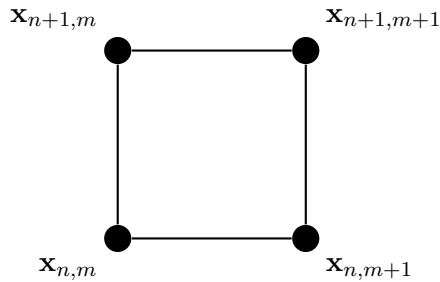
\begin{figure}[bthp]
   \centering
   \begin{tikzpicture}
       \node (x1) at (0,0) [circle,fill,label=-135:$\vec{x}_{n,m}$] {};
       \node (x4) at (0,2.5) [circle,fill,label=135:$\vec{x}_{n+1,m}$] {};
       \node (x2) at (2.5,0) [circle,fill,label=-45:$\vec{x}_{n,m+1}$] {};
       \node (x3) at (2.5,2.5) [circle,fill,label=45:$\vec{x}_{n+1,m+1}$] {};
      \draw [thick] (x2) to (x1);
      \draw [thick] (x4) to (x3);
      \draw [thick] (x3) to (x2);
      \draw [thick] (x1) to (x4);
   \end{tikzpicture} 
   \caption{A quad-graph.}
\label{fig:quad}
\end{figure}

In this paper I will address the problem of the calculation of the
algebraic entropy for \emph{systems of quad equations}.  That is,
I will consider systems of four-point relations for an unknown field
$\vec{x}\colon\Z^{2}\to\Cp^{M}$ of the form:
\begin{equation}
    \vec{Q}(\vec{x}_{l,m},\vec{x}_{l+1,m},\vec{x}_{l,m+1},\vec{x}_{l+1,m+1})=
    \vec{0},
    \label{eq:quadsys}
\end{equation}
and I will address their growth properties in the sense of Arnol'd
\cite{Arnold1990}. Equations of this form appeared seldom in the
literature, see e.g. \cite{KonstantinouRizos_etal2015, Hay2009,Hay2011,
Kels2019Z, Nijhoff_etal1992, Nijhoff1996Dorfmann, XenitidisNijhoff2012}.
More recently, the interest in these models appears to be increasing, see
\cite{BradyXenitidis2022}, where their generalised (continuous)
symmetries are discussed. The name \emph{quad equations} originates from
the fact that the four-point relation is defined on a square-like shape,
see \Cref{fig:quad}.

I started studying the growth properties of difference equations during my
Ph.D. at Universit\`a degli Studi Roma Tre under the supervision of Prof.
Decio Levi, and this paper is dedicated to his memory.  Some of the
computations contained in this paper date back to the time I was a Ph.D.\
student but were not published for several reasons, while others are new
and were carried out explicitly for this occasion. For these reasons this
paper is written in first person singular and it will have two
introductions: one, more personal, about my relationship with Decio Levi,
which lasted until the very end of his life, and a scientific one, where I
will address the background of the problem and the structure of this paper.

\subsection{My relationship with Decio Levi}

Before starting my Ph.D. in Mathematics at Universit\`a degli Studi Roma
Tre I knew a little about Decio Levi's work after the early 90s . During
the preparation of my Bachelor Thesis at Universit\`a degli Studi di
Perugia \cite{Gubbiotti2010Thesis} I studied several of the very
interesting papers produced by Decio and Pavel Winternitz
\cite{LeviMenyukWinternitz1991, DavidKamranLeviWinternitz1985a,
DavidKamranLeviWinternitz1985b} regarding the optimal Lie subalgebras (see
\cite[\S 3.3]{Olver1986}) of infinite dimensional Lie algebras. I had a
strong interest in integrable systems since I attended a seminar of Mark J.
Ablowitz at Universit\`a degli Studi di Perugia. Moreover, I just came from
writing a Master's Thesis at Universit\`a degli Studi di Trieste under the
supervision of Tamara Grava on the asymptotic properties of NLS equations
with piece-wise constant initial data \cite{DiFranco2005}, and I was
actively working with Maria Clara Nucci on Noether symmetry preserving
quantisation \cite{GubNucLienI}. On the other hand, I had little or no idea
of discrete systems. During my very first international conference, the
Twenty Years of Journal of Nonlinear Mathematical Physics conference in
Nordfjordeid, I think I skipped almost all sessions about discrete systems
because I understood almost nothing about that subject at that time.

So, with little or no knowledge of the work of Decio after the early 90s we
had our first email exchange, where I asked for an idea of a research
project to present to the Ph.D. admission test. I mentioned as a reference
Maria Clara Nucci, and after asking her about me, Decio proposed me two
topics:
\begin{itemize}
    \item the problem of symmetry preserving discretisation
        \cite{WinternitzASIDE};
    \item the study of the new quad equations classified by R. Boll in his
        Ph.D. thesis \cite{Boll2011,Boll2012a,Boll2012b}.
\end{itemize}
After some study of the references he provided I opted for the second one.

Our first real life encounter was after I already won a Ph.D. position at
Universit\`a degli Studi Roma Tre. I went to bring the documents needed for
the enrollment, and then I went to Physics Building in Via della Vasca
Navale 84, got lost in the maze of the corridors and finally found his
office. At that time Ravil Islamovich Yamilov was there. Our first meeting
was very pleasant. To give some additional context, in those years
integrable systems were a very well developed topic at Universit\`a degli
Studi Roma Tre: the same year I got admitted as a Ph.D.\ student in
Mathematics, Danilo Latini got admitted in Physics, and choose as
supervisor Orlando Ragnisco, while Christian Scimiterna and Federico Zullo
were postdocs, supported by the very large PRIN ``Geometric and Analytic
theory of Hamiltonian systems in finite and infinite dimensions''.  There
was the local INFN unit of the project ``Mathematical Methods of NonLinear
Physics'', which is the heir of the original Italian school of
integrability led by Francesco Calogero, Giulio Soliani, and many others.
In short the milieu was very fertile for developing ideas and growing as a
scientist.

After almost one year of classes and study we started the real work.
Since, as I told before, I was completely new to discrete integrable
systems I started by reading \cite{HietarintaBook}. My original task was to
compute the generalised symmetries \cite{LeviYamilovASIDE} of the Boll's
new equations on the quad-graph. To cut a long story short, the generalised
symmetry test for integrability of quad equations states that a
quad-equation is integrable if its generalised symmetries are themselves
integrable equations as flows, i.e.~they are integrable
differential-difference (or semi-discrete) equations
\cite{LeviYamilov2009}. At first I was sort of puzzled by Boll's
equations, and having studied the algebraic entropy test, which looked like
magic to me back then (sometimes it still does even nowadays) I proposed
to Decio to check first the equations with algebraic entropy.  At that time
Christian Scimiterna was post-doc at Universit\`a degli Studi Roma Tre and
suggested that I ask Mike Hay, who was also there with an INFN fellowship,
since they recently faced a similar problem in
\cite{ScimiternaHayLevi2014}. Decio was supportive of this idea, even
though his main focus was far away from the ideas of algebraic entropy.

After a few months of studying and coding, I came up with a suite of
programs to compute the growth of degrees of quad equations and analyse
their behaviour. The day of reckoning came and I got the unexpected result
that the equations we were considering (that Christian carefully checked to
be independent and ``realisable'' in Boll's 3D cubes configurations)
possessed linear growth. Even more surprising while looking at pages and
pages of output from my programs I found out that contrary to what was
assumed in several papers on the topic, e.g.~\cite{Viallet2006,
Tremblay2001}, the degrees along the diagonal were different. There were
several days we spent in doubt about these results, which were resolved
after I sent an email to Claude Viallet, where he confirmed that the result
was plausible. Thus, my first paper with Decio and Christian was born
\cite{GSL_general}, even though it was not the first to be published. We
further elaborated some of these ideas in \cite{GSL_Gallipoli15}, where upon
the observation that linear(isable) equations cannot possess infinitely
many conservations laws, Decio suggested to me and Christian to prove that
the Lax pair of the H1\textsubscript{t} equations, the simplest equations
we were considering, obtained by the Consistency Around the Cube procedure
was in fact \emph{fake} \cite{CalogeroNucci1991, HayButler2013,
HayButler2015}, i.e. it was just a way of rewriting the equation, and it
was not giving any additional information.

During the rest of my Ph.D. I spent most of my time trying to understand
more about the results we gathered in \cite{GSL_general}. After I first
presented our work together at the Physics and Mathematics of Nonlinear
Phenomena 2015, we went back to the original idea and computed the
generalised symmetries of Boll's equations \cite{GSL_symmetries}. When
Decio unexpectedly remembered his first work with Ravil
\cite{LeviYamilov1997} we realised that the flows of all the non-autonomous
generalised symmetries we computed were actually particular cases of an
equation introduced by Decio and Ravil themselves! That observation, and a
comparison with the works of Xenitidis and
Papageorgiou~\cite{Xenitidis2009} and Xenitidis~\cite{Xenitidis2009proc},
led us to build a two-periodic extension of the Q\textsubscript{V} equation
\cite{Viallet2009} in \cite{GSL_QV}. This is just one of the many cases
when Decio's experience and cunning were fundamental.

In writing \cite{GSL_symmetries} the help of Christian in this task was
again fundamental, as he noticed that the trapezoidal
H1\textsubscript{$\varepsilon$} equation had a very peculiar property: it
admitted generalised symmetries depending on \emph{arbitrary functions}.
This led us to the concept of \emph{Darboux integrability} for quad
equations as exposed in \cite{AdlerStartsev1999}. This notion was the
subject of the last year of my Ph.D., and of the last paper I wrote
together with Decio and Christian \cite{GSL_Pavel}. Therein, we proposed a
conjecture, strongly advocated by Decio, that all quad equations admitting
generalised symmetries depending on arbitrary functions were in fact
Darboux integrable. This conjecture was shortly afterwards proved true by
Startsev \cite{Startsev2016}.  Also during the last year of my Ph.D. Ravil
was visiting Universit\`a degli Studi di Roma Tre. He had won an INdAM
visiting position, and he was supposed to give a Ph.D.\ course. However, it
turned out that I was the only student. So, after a few classic lectures,
upon Decio's idea we started developing our ideas on the Darboux
integrability of Boll's equations instead of doing lectures. The result of
these anomalous lectures was two papers \cite{GSY_DarbouxI,GSY_DarbouxII},
where we solved the mystery of Boll's equations. Indeed, we explained the
linear growth by showing that the trapezoidal H4 and the H6 equations are
Darboux integrable. Moreover, we proved that they are exactly solvable, up
to some discrete Riccati equations. My collaboration with Ravil continued
in the later years, see \cite{GGY_autom}, and it lasted until Ravil's
premature demise.

After that, I got my Ph.D. and I moved to Australia. Indeed, in the last
part of 2016, I won a post-doc position at the University of Sydney in the
Integrable and Nearly Integrable Systems group led by Nalini Joshi. In some
sense, I also owe this possibility to Decio, because my stronger connection
to Australia, started when following his idea I went to Australia in 2015,
touring Melbourne, Brisbane, and Sydney. In time, our encounters become
sporadic: we met once in Sydney in 2017, once in Fukuoka at the SIDE13
conference in 2018, and at the special session for his 71st birthday in
Montreal. On the few occasions when I came back to Universit\`a degli Studi
Roma Tre after I got my Ph.D., I went back to his study (converted to a
joint study for him and Orlando), he was not there, because he preferred
working from home.  However, we frequently exchanged emails. A few times I
tried to involve him in doing some additional work together, but
(unfortunately) he was always turning my proposals down, yet pushing me to
follow my ideas by myself.  At that point his main interests were the
conditional symmetry preserving schemes for PDEs \cite{Levietal2021}, and
his long time project with Pavel and Ravil: the book
``\citetitle{LeviWinternitzYamilov2022Book}''
\cite{LeviWinternitzYamilov2022Book} which was finally published in the
last quarter of 2022, unfortunately after the demise of all its authors. 

I always resorted to Decio when I was in need of a suggestion about my
career and my scientific results. Decio was always very calm, and he was
the perfect balance to my illtemper. As a supervisor, Decio was really
good. It was clear that he was doing the best for both of the parties, the
students and the supervisor, helping you to grow and become independent,
not just some mindless calculator. My very first sole author work, the
review \cite{GubbiottiASIDE16}, followed his invitation to give a four
hours lecture at ASIDE, the summer school right before SIDE13 that he was
organising  with Pavel in Sainte Ad\`ele. During my Ph.D.\ I was also left
free to have my own scientific collaborations, for instance my ongoing
collaboration with Maria Clara Nucci, which resulted in a few more works
\cite{GubNucLienII,GubNucCone,GubNucSupint1}, or my new collaboration with
Davide Chiuchi\`u \cite{ChiuchiuGubbiotti01,ChiuchiuGubbiotti02} on a
completely different topic than the rest of my Ph.D.\ thesis.

\emph{Losing Decio, I did not just lose my supervisor and mentor. I lost
one of the nicest persons you can find on earth, and I will always miss him
and his soft-spoken suggestions.}

\subsection{Statement of the problem and outline of the paper}

Analysing algorithms through their complexity is an old topic in numerical
analysis. Heuristically, the idea is that algorithms performing better are
those whose computational time is sub-exponential in the iteration process.
The goal is then to establish the long term behaviour of the recurrence
defined by the algorithm \emph{without having to compute the complete
recurrence or knowing how to express it in closed form}. Clearly, computing
infinitely many terms is impossible, and the examples that one can exactly
solve are very few. Moreover, having a solution does not necessarily mean
that the system is asymptotically well-behaved, see for instance the
solvable case of the logistic map \cite{MayNature1976,Schroder1870}.

%This simple example shows how examining the iterates of a recurrence
%relation can be a good way to extract information about integrability even
%if we cannot solve the equation explicitly. However in more complicate
%examples it is usually impossible to calculate explicitly these iterates
%by hand or even with any state-of-the-art formal calculus software, simply
%because the expressions one should manipulate are rational fractions
%of increasing degree of the various initial conditions.  The complexity
%and size of the calculation make it impossible to  calculate the iterates.

In this spirit, in \cite{Arnold1990} Arnol'd defined the complexity of
intersections for diffeomorphisms. In fact the idea is rather simple: let
us assume we are given a compact smooth manifold $\mathcal{X}$, a
diffeomorphism $\varphi\in\mathcal{C}^{\infty}(\mathcal{X})$ and two compact
smooth submanifolds $\mathcal{Y}$, $\mathcal{Z}$, such that $\dim
\mathcal{Y}+\dim \mathcal{Z} = \dim \mathcal{X}$. Then a natural way to
estimate the complexity upon iteration of the diffeomorphism $\varphi$ is
to consider the cardinality of the intersection of the successive images of
$\mathcal{Y}$ with $\mathcal{Z}$:
\begin{equation}
    N_{k} = \abs{\varphi^{(k)}(\mathcal{Y})\cap \mathcal{Z}}.
    \label{eq:arnold}
\end{equation}
The ``generic'' growth for such intersections grows exponentially and in
some pathological cases can even be super-exponential.

What was observed is that there are some particular maps (which will be
later called ``integrable'') that are not as complex as generic ones
\cite{Veselov1992,FalquiViallet1993,HietarintaViallet1997,Diller1996}.
These maps had to be birational maps of complex projective spaces,  because
the drop in complexity relies on the ability of the map to enter (and
possibly exit) the singularities. Then Bellon and Viallet introduced the
notion of \emph{algebraic entropy} \cite{BellonViallet1999} to have an
invariant measure of growth of birational maps of the complex spaces.
Almost at the same time \cite{Russakovskii1997} introduced the analogous
notion of \emph{dynamical degree}. Since then the development of the theory
of algebraic entropy has been a thriving research topic, both for the
discrete integrable systems community and for the algebraic dynamics one,
see for instance the reviews
\cite{GrammaticosHalburdRamaniViallet2009,GubbiottiASIDE16}.

In particular, over the years the theory of algebraic entropy has been
extended beyond birational maps of complex projective spaces, to several
infinite-dimensional cases. For instance in \cite{Tremblay2001,Viallet2006}
the method was developed in the case of scalar quad equations, while in
\cite{GSL_general} a slight generalisation was proposed. For instance, a
search for integrable quad equations using factorisation was used in
\cite{HietarintaViallet2007}. Later the concept was extended to
semi-discrete systems: in \cite{DemskoyViallet2012} for
differential-difference equations, and in \cite{viallet2014} for
differential-delay equations. Lattice equations not of quad type have been
considered more recently
\cite{GubKelsEntropy,GKVHexEntropy,Hietarinta2023Bous}, while
quad equations themselves are still the source of new findings
\cite{Hietarintaetal2019}.

Here, I will discuss the concept of algebraic entropy for systems of
quad equations, i.e. pure difference systems of the form
\eqref{eq:quadsys}. Then, the structure of the paper is the following: In
\Cref{sec:ae} I will present the technique of calculation of the degree
sequences for systems of quad equations, and I will recall how to estimate
asymptotic growth. My treatment will follow closely the construction that
is done usually in the scalar case, see
\cite{Viallet2006,Tremblay2001,GSL_general}, with the peculiar difference
that multilinear systems might not have a well defined birational evolution
in all directions. In \Cref{sec:examples} I will present several examples
of systems that pass the algebraic entropy test. When possible, I will
discuss the similarities and differences with the scalar case. Finally, in
\Cref{sec:concl} I will discuss the results obtained.

\section{Algebraic entropy for systems of quad equations with staircase initial conditions}
\label{sec:ae}

In this paper I will consider the problem of computing the algebraic entropy of
systems of quad equations using initial conditions of a specific form, i.e.\
the so-called \emph{staircase initial conditions}. 

In the case of scalar quad equations
\begin{equation}
    Q(x_{l,m},x_{l+1,m},x_{l,m+1},x_{l+1,m+1})=0,
    \label{eq:scalquad}
\end{equation}
the multilinearity requirement ensures that the evolution is well-defined as a
rational map in all the directions of evolution. That is, solving a quad equation
with respect to one of the four corners yields a rational map. The evolution is
then birational in the opposite direction with respect to the principal
diagonal, i.e.\ solving with respect to the pair $(x_{l+1,m},x_{l,m+1})$ and
$(x_{l,m},x_{l+1,m+1})$.

In the case of systems, this is no longer true. The reason is twofold: first I
will not assume that all equations in the system \eqref{eq:quadsys} depend on all
four vertices. Then, a first condition to have solvability in one of the four
directions is:
\begin{equation}
    \rank \frac{\partial \vec{Q}}{\partial \vec{x}_{l+1,m+j}} = M,
    \quad
    i,j\in\Set{0,1},
    \label{eq:jacquadsys}
\end{equation}
since it allows to use the inverse function theorem.  However, this is not
enough: if a system of multilinear equations is solvable it does not
necessarily imply that the solution is \emph{rational}. A simple example is
the following:
\begin{subequations}
    \begin{align}
        Q_{1} &=
        x_1 y_1 y_2 x_3-x_2 y_3 y_4 x_4,        
        \\
        Q_{2} &=
        x_1 y_2+y_3 x_4+y_1 x_2+x_3 y_4,
    \end{align}
    \label{eq:exsimpl}
\end{subequations}
were for simplicity I used the variables $x_{i}$, $y_{i}$ with $i=1,2,3,4$.
Then we have:
\begin{equation}
    \rank \frac{\partial (Q_{1},Q_{2})}{\partial (x_{1},y_{1})} = 
    \rank \frac{\partial (Q_{1},Q_{2})}{\partial (x_{3},y_{3})} = 
    2,
    \label{eq:jacquadsysex}
\end{equation}
but in the first case the solutions are algebraic:
\begin{subequations}
    \begin{align}
        x_1 = \frac{1}{2} \frac{-x_3^2 y_4-x_3 x_4 y_3\mp\sqrt{-4 x_2^2 x_3 x_4 y_3 y_4+x_3^4 y_4^2+2 x_3^3 x_4 y_3 y_4+x_3^2 x_4^2 y_3^2}}{x_3 y_2}, 
        \\
        y_1 = \frac{1}{2} \frac{-x_3^2 y_4-x_3 x_4 y_3\pm \sqrt{-4 x_2^2 x_3 x_4 y_3 y_4+x_3^4 y_4^2+2 x_3^3 x_4 y_3 y_4+x_3^2 x_4^2 y_3^2}}{x_2 x_3}
    \end{align}
    \label{eq:algsols}
\end{subequations}
while in the second are rational:
\begin{equation}
    x_3 = -x_2 y_4\frac{x_1 y_2+x_2 y_1}{x_1 y_1 y_2+x_2 y_4^2},  
    \quad
    y_3 = - \frac{x_1 y_1 y_2}{x_4} \frac{x_1 y_2+x_2 y_1}{x_1 y_1 y_2+x_2 y_4^2}.
    \label{eq:ratsols}
\end{equation}
When solutions are rational, I will say that the direction of evolution
is \emph{admissible}, while the solutions are not rational, I will say 
that the direction of evolution is \emph{not admissible}.

%In the case of quad equations the situation is more complicate than in the
%case of one-dimensional equations.  First of all in the one-dimensional
%case one have to worry only about the evolution in two opposite directions
%(this was the meaning of the bi-rationality condition). In the two
%dimensional case we have to worry about four possible directions of
%evolution corresponding to the four ways I can solve the quad equation.
%In general, initial conditions can be given along straight lines in the
%four direction.  
%However usually is preferred to give initial conditions
%on \emph{staircase} configurations, which will truly correspond to
%initial conditions.

\begin{figure}[hbt]
    \centering
    \begin{tikzpicture}[scale=0.5]
         \draw[style=help lines,dashed] (0,0) grid[step=1cm] (15,15);
         \draw[thick] (5,0)--(5,2)--(4,2)--(4,4)--(3,4)--(3,6)--
            (2,6)--(2,8)--(1,8)--(1,10)--(0,10)--(0,12);
         \node[right] at (0,12) {(1)};
         \draw[thick, red] (15,8)--(15,9)--(14,9)--(14,10)--(13,10)--(13,11)--
            (12,11)--(12,12)--(11,12)--(11,13)--(10,13)--(10,14)--(9,14)--(9,15)--(8,15);
        \node[below, red] at (8,15) {(2)};
        \draw[thick, blue] (0,5)--(4,5)--(4,6)--(5,6)--(7,6)--(7,10)--(8,10)--(8,11)--(13,11)--(13,15)--(15,15);
        \node[below left, blue] at (15,15) {(3)};
        \draw[thick, teal] (15,2)--(13,2)--(13,4)--(11,4)--(11,6)--(13,6)--(13,7)--(10,7)--(8,7)--
            (8,9)--(11,9)--(11,12)--(8,12)--(7,12)--(7,15)--(5,15);
        \node[below, teal] at (5,15) {(4)};
     \end{tikzpicture}
    \caption{Regular and non-regular staircases.}
    \label{fig:staircases}
\end{figure}
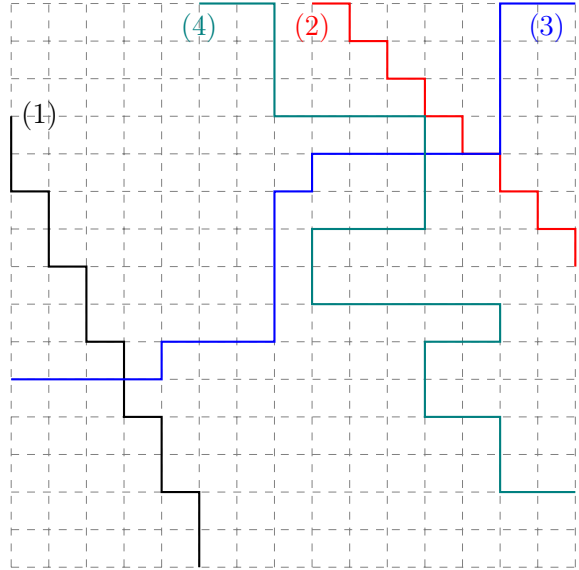

Then, if the condition of admissibility is satisfied, the evolution is possible
on a staircase-like arrangement of initial values is possible in the
quadrilateral lattice. In such a case the treatment of the problem will follow
closely the scalar case (see e.g.~\cite{Viallet2006, Tremblay2001,
GubbiottiASIDE16, GrammaticosHalburdRamaniViallet2009, GSL_general}), with some
relevant differences I will point out in the discussion. The first thing to rule
out is overhang in the initial data configurations, since these could lead to a
contradiction, giving more than one way to calculate the same value for the
dependent variable.  The staircases need to go from $(n=-\infty, m=-\infty)$ to
$(n=\infty, m=\infty)$, or from $(n=-\infty, m=+\infty)$ to $(n=\infty,
m=-\infty)$ because the space of initial conditions is infinite. I will
restrict the present discussion to regular diagonals which are staircases with
steps of \emph{constant} horizontal length, and \emph{constant} vertical
height.  Non-regular staircases were considered in the scalar case
in~\cite{Hietarintaetal2019}, raising several interesting questions, which as
far as I know are not completely resolved yet.  \Cref{fig:staircases} shows
four diagonals. The ones labeled $(1)$ and $(2)$ are regular. The one labeled
$(3)$ would be acceptable, but non-regular, so I will not consider it. Line
$(4)$ is excluded since it may lead to incompatibilities.  In the scalar case,
given a line of initial conditions, it is possible to calculate the values
$x_{l,m}$ on all points of the two-dimensional lattice. In the system case,
this is possible if both directions, are admissible.  With the examples I will
consider in \Cref{sec:examples} this is not always the case, \emph{yet the
growth might still be sub-exponential}.

The crucial point of using regular diagonals, is that \emph{if one wants the
evolution on a finite number of iterations, only a diagonal of initial
conditions of finite extent is needed}.  Indeed, let $N$ be a positive integer,
and each pair of relative integers ${[\lambda_1,\lambda_2]}$, I will denote by
$\Delta_{[\lambda_1,\lambda_2]}^{(N)}\subset \Z^{2}$ a regular diagonal
consisting of $N$ steps, each having horizontal size $l_1 = \vert \lambda_1
\vert$, height $l_2 = \vert \lambda_2 \vert$, and going in the direction of
positive (resp.\ negative) $n_k$, if $\lambda_k > 0$ (resp.  $\lambda_k < 0$),
for $k=1,2$. See \Cref{fig:regulardiagonals}.

\begin{figure}[hbt]
    \centering
    \begin{tikzpicture}[scale=0.5]
         %\draw[style=help lines,dashed] (0,0) grid[step=1cm] (14,14);
         \draw[style=help lines,dashed] (0,0) grid[step=1cm] (18,8);
         \draw[thick] (0,0)--(1,0)--(1,1)--(2,1)--(2,2)--(3,2)--(3,3);
         \node[below] at (0,0) {$\Delta_{[1,1]}^{(3)}$};
         \draw[thick, red] (3,8)--(3,5)--(4,5)--(4,2)--(5,2);
         \node[above, red] at (3,8) {$\Delta^{(2)}_{[-1,3]}$};
         \draw[thick, blue] (6,0)--(6,2)--(9,2)--(9,4)--(12,4)--(12,6)--(15,6)--
            (15,8)--(18,8);
            \node[above, blue] at (18,8) {$\Delta^{(4)}_{[-3,-2]}$};
        \draw[thick, teal] (12,3)--(14,3)--(14,2)--(16,2)--(16,1)--(18,1)--(18,0);
        \node[below, teal] at (12,3) {$\Delta_{[2,-1}^{(3)}$};
     \end{tikzpicture}
    \caption{Varius kinds of restricted initial conditions.}
    \label{fig:regulardiagonals}
\end{figure}

Suppose one fixes the initial conditions
$\Delta_{[\lambda_1,\lambda_2]}^{(N)}$. Then, one can calculate the values
$\vec{x}_{l,m}$ over a rectangle of size $(N l_1+1) \times (N l_2 +1)$. The
diagonal cuts the rectangle in two halves: one of them uses all initial
values.  Assuming the associated direction of evolution is admissible for
the system~\eqref{eq:quadsys} one can then use those initial values to
compute the evolution on the part using all the initial values, see
\Cref{fig:range}. We call this set of indices
$\Theta_{[\lambda_{1},\lambda_{2}]}^{(N)}$, the \emph{range} of
$\Delta_{[\lambda_1,\lambda_2]}^{(N)}$

\begin{figure}[hbt]
    \centering
    \begin{tikzpicture}[scale=0.5,every node/.style={draw,shape=circle,fill=black, scale=0.3}]
        \draw[thick] (0,0)--(-2,0)--(-2,1)--(-4,1)--(-4,2)--(-6,2)--(-6,3);
        \node at (0,0) {}; 
        \node at (-1,0) {};
        \node at (-2,0) {}; 
        \node at (-2,1) {}; 
        \node at (-3,1) {}; 
        \node at (-4,1) {}; 
        \node at (-4,2) {}; 
        \node at (-5,2) {}; 
        \node at (-6,2) {}; 
        \node at (-6,3) {}; 
        \draw[dashed]  (0,0)--(0,3)--(-6,3);
        \draw[dashed]  (-1,0)--(-1,3);
        \draw[dashed]  (-2,1)--(-2,3);
        \draw[dashed]  (-3,1)--(-3,3);
        \draw[dashed]  (-4,2)--(-4,3);
        \draw[dashed]  (-5,2)--(-5,3);
        \draw[dashed] (-2,1)--(0,1);
        \draw[dashed] (-4,2)--(0,2);
        \node at (0,1) {};
        \node at (0,2) {};
        \node at (0,3) {};
        \node at (-1,1) {};
        \node at (-1,2) {};
        \node at (-1,3) {};
        \node at (-2,2) {};
        \node at (-2,3) {};
        \node at (-3,2) {};
        \node at (-3,3) {};
        \node at (-4,3) {};
        \node at (-5,3) {};
     \end{tikzpicture}
     \caption{The range for the initial conditions $\Delta_{[-2,1]}^{(3)}$.}
    \label{fig:range}
\end{figure}

So, on some restricted diagonal $\Delta_{[\lambda_1,\lambda_2]}^{(N)}$, such
that the direction using all the initial values is admissible, the iteration is
defined. Then, by the previous discussion:
\begin{equation}
    \abs{\Delta_{[\lambda_1,\lambda_2]}^{(N)}} = N( l_1 + l_2)+1=:\mathcal{N}.
    \label{eq:Nin}
\end{equation}
So, the total number of initial conditions needed is $\mathcal{N}\cdot M$,
which in principle are an arbitrary vector in $\Cp^{\mathcal{N}\cdot M}$.
However, as usual, it is better to consider the initial values in a
\emph{compactification} of the complex space. My choice for this paper is to
consider as compactification the space:
\begin{equation}
    \mathcal{K}_{\mathcal{N},M} = \bigl( \Pj^{\mathcal{N}} \bigr)^{\times M}.
    \label{eq:spaceK}
\end{equation}
For instance other possibile compactifications are $\Pj^{\mathcal{N}\cdot
M}$ and $\bigl( \Pj^{1} \bigr)^{\mathcal{N}\cdot M}$. However, the choice
of the compactification will not change the final result.  My choice of the
space $\mathcal{K}_{\mathcal{N},M}$ is due to the fact that I want to keep
track of every component of $\vec{x}_{l,m}$ as separate entities.  In
practice, to simplify the computations, it is usually better to consider
the intial data as lying on a \emph{line} in $\mathcal{K}_{\mathcal{N},M}$,
i.e.  it is possible to write them in the following form:
\begin{equation}
    x_{i,j}^{k}
    =
    \frac{\alpha_{i,j}^{k}t_{0}+\beta_{i,j}^{k}t_{1}}{\alpha_{0}^{k}t_{0}+\beta_{0}^{k}t_{1}},
    \quad
    (i,j)\in \Delta_{[\lambda_1,\lambda_2]}^{(N)},
    [t_{0}:t_{1}]\in\Pj^{1},
    k=1,\dots,M,
    \label{eq:iniexpl}
\end{equation}
where $\vec{x}_{i,j}= \bigl( x_{i,j}^{1},\dots,x_{i,j}^{M}\bigr)$ and the
coefficients $\alpha_{i,j}^{k}$, $\beta_{i,j}^{k}$, $\alpha_{0}^{k}$, and
$\beta_{0}^{k}$ are (fixed) integers.  Then, evaluating the equation on the
points reachable from $\Delta_{[\lambda_{1},\lambda_{2}]^{(N)}}$ we obtain
a two-dimensional sequence of degrees $\Set{\vec{d}_{l,m}}_{(l,m)\in
\Theta_{[\lambda_{1},\lambda_{2}]}^{(N)}}$.  Here the vector
$\vec{d}_{l,m}=(d_{i,l,m},\dots,d_{M,l,m})^{T}$, means that to any
component of the field $\vec{x}_{l,m}$ we associate a different degree, and
we whish to keep track of it through the evolution. When the number of
fields is small I will use letters  rather than numbering, e.g. three
fields $\vec{x}_{l,m}=(x_{l,m},y_{l,m},z_{l,m}^{T})$. Correspondingly I
will denote the degrees by letters, e.g. continuing the previous examples
the vector of degrees will be
$\vec{x}_{l,m}=(d_{x,l,m},d_{y,l,m},d_{z,l,m}^{T})$.

\begin{figure}[hbt]
    \centering
    \begin{tikzpicture}[scale=0.5]
         \draw[style=help lines,dashed] (0,0) grid[step=1cm] (14,14);
         \draw[thick] (0,9)--(0,10)--(1,10)--(1,11)--(2,11)--(2,12)--(3,12)--(3,13)--(4,13)--(4,14)--(5,14);
         \node[above right] at (3+1/2,11-1/2) {$\Delta_{-,-}$};
         \draw[thick,->] (2+1/2,12-1/2)--(3+1/2,11-1/2); 
         \draw[thick,red] (0,7)--(0,6)--(1,6)--(1,5)--(2,5)--(2,4)--(3,4)--(3,3)--(4,3)--(4,2)--(5,2)--(5,1)--(6,1)--(6,0)--(7,0);
         \node[above,red] at (5-1/2,4+1/2) {$\Delta_{+,-}$};
         \draw[thick,->,red] (4-1/2,3+1/2)--(5-1/2,4+1/2); 
         \draw[thick,blue] (8,0)--(9,0)--(9,1)--(10,1)--(10,2)--(11,2)--(11,3)--(12,3)--(12,4)--(13,4)--(13,5)--(14,5)--(14,6);
         \node[above,blue] at (11-1/2,4+1/2) {$\Delta_{+,+}$};
         \draw[thick,->,blue] (12-1/2,3+1/2)--(11-1/2,4+1/2); 
         \draw[thick,teal] (14,7)--(14,8)--(13,8)--(13,9)--(12,9)--(12,10)--(11,10)--(11,11)--(10,11)--(10,12)--(9,12)--(9,13)--(8,13)--(8,14)--(7,14);
         \node[below,teal] at (11-1/2-1,11-1/2-1) {$\Delta_{-,+}$};
         \draw[thick,->,teal] (11-1/2,11-1/2)--(11-1/2-1,11-1/2-1); 
     \end{tikzpicture}
    \caption{The four principal diagonals.}
    \label{fig:princdiag}
\end{figure}
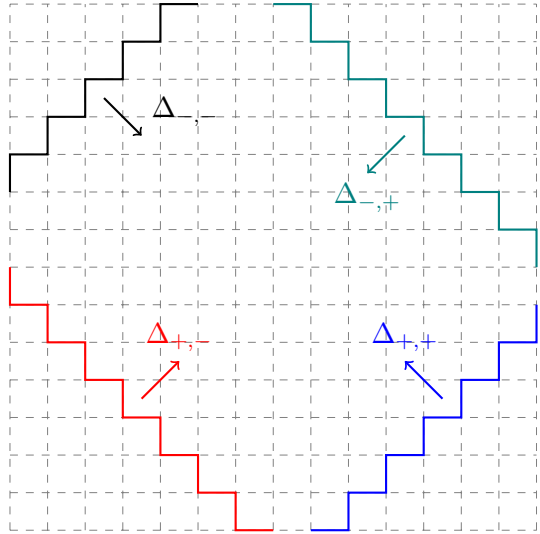

Among all possible restricted staircases, a special r\^ole is played by the
simplest ones, i.e. the restricted diagonals $\Delta_{[\pm 1, \pm
1]}^{(N)}$. Due to their importance I denote them by the special notations
$\Delta_{++}^{(N)}$, $\Delta_{+-}^{(N)}$, $\Delta_{-+}^{(N)}$ and
$\Delta_{--}^{(N)}$, and name them \emph{the fundamental diagonals}. As
usual, the upper index $(N)$ is omitted for infinite lines. The four
fundamental diagonals are showed in Figure \ref{fig:princdiag}.

Explicitly, a degree sequence on $\Delta_{+-}^{(N)}$ has the form:
\begin{equation}
    \begin{array}{cccccc}
        1 & \vec{d}_{1,N-1} & \vec{d}_{2,N-2} & \dots & \vec{d}_{N-1,2} 
        & \vec{d}_{N,1}
        \\
        1 & 1 & \vec{d}_{1,N-2} & \vec{d}_{2}^{(3)} & \dots & \vec{d}_{N-1,1}
        \\
        & 1 & 1 & \vec{d}_{1,3} & \vec{d}_{2,2} & \dots
        \\
        & & 1 & 1 & \vec{d}_{1,2} & \vec{d}_{2,1}
        \\
        & & & 1 & 1 & \vec{d}_{1,1}
        \\
        & & & & 1 & 1
    \end{array}
    \label{eq:ourdegrees}
\end{equation}
In \Cref{eq:ourdegrees} $1$ simply means that all sequence starts with a
term of homogeneous degree $1$, as in \Cref{eq:iniexpl}. Note that we
choose the indexing in a way that the first index represents the index of
iteration. Moreover, if needed, to underline the direction in the sequence
of degrees we will denote by $\vec{d}_{l,m}^{\pm \pm}$ the sequence of
degrees on corresponding to $\Delta_{\pm , \pm }$. Then, for every fixed $m$
we define the sequence
\begin{equation}
    1,\vec{d}_{1,m}^{\pm\pm},\vec{d}_{2,m}^{\pm\pm},
    \vec{d}_{3,m}^{\pm\pm},\vec{d}_{4,m}^{\pm\pm},\dots,
    \label{vialletseq}
\end{equation}
Usually, these sequences are just the same sequence shifted by one. In such
a case we simply drop the unused index $m$, and write the sequence as
$\Set{\vec{d}_{l}^{\pm\pm}}_{l=0}^{N}$.  In the known cases where there is more
than one sequence living on the lattice of \Cref{eq:ourdegrees}, the
sequences are repeating periodically, see \cite{GSL_general}. In the
examples presented in this paper there will be only one sequence for
admissible directions. To simplify our discussion of the examples in
\Cref{sec:examples} we introduce now some notations and definitions.

\begin{notation}
    I will denote a generic element of the set $\Set{\pm}^{2}$ by $\delta$,
    and consequently the degree sequence in the direction
    $\delta\in\Set{\pm}^{2}$ will be denoted by $\Set{\vec{d}_{l,m}^{\delta}}_{(l,m)\in
    \Theta^{(N)}_{\delta}}$. 
\end{notation}

\begin{definition}
    Let us assume we are given a system of quad-equation
    \eqref{eq:quadsys}.  Then:
    \begin{itemize}
        \item the system is \emph{isotropic} if for all admissible 
            directions the sequence $\vec{d}_{l,m}^{\delta}$ is the same;
        \item the system is \emph{strongly isotropic} if for all admissible 
            directions the sequences $d_{i,l,m}^{\delta}$, $i=1,\dots,M$
            are the same;
        \item the system is \emph{permutationally isotropic} if for
            all pairs of admissible directions $\delta_{1}$ and
            $\delta_{2}$ there exists a permutation
            $\sigma\in \mathcal{S}_{M}$ such that:
            \begin{equation}
                \sigma(\vec{d}_{l,m}^{\delta_{2}}) := 
                (d_{\sigma(1),l,m}^{\delta_{2}},\dots,d_{\sigma(M),l,m}^{\delta_{2}})\equiv
                \vec{d}_{l,m}^{\delta_{1}}.
                \label{eq:permisot}
            \end{equation}
    \end{itemize}
    \label{def:isotropy}
\end{definition}

\begin{remark}
    Note that the system being isotropic means that \emph{the degree growth
    does not depend by the (admissible) direction}, being strongly
    isotropic means that the \emph{the degree growth
    does not depend by the (admissible) direction \emph{and} the field},
    and finally that being permutationally isotropic means \emph{the degree growth
    does is only permuted by the direction}. In all cases, the system of
    quad equations must have \emph{more than one} admissible direction.
    \label{rem:isot}
\end{remark}

%We shall call the sequence
%the \emph{ principal sequence} of growth. A sequence as 
%\begin{equation}
%    1, d_{1,\pm\pm}^{(l)}, d_{2,\pm\pm}^{(l)}, d_{3,\pm\pm}^{(l)}, 
%        d_{4,\pm\pm}^{(l)},\ldots
%    \label{eq:ourseq}
%\end{equation}
%with $l=2$ will be a \emph{ secondary sequence} of growth, for 
%$l=3$ a \emph{ third}, and so on. 

Then define the \emph{fundamental algebraic entropies} of the
lattice equation by:
\begin{equation}
    S_{i,m}^{\delta} = \lim_{l\to\infty}\frac{1}{l}\log
    d_{i,l,m}^{\delta}, \quad i=1,\dots,M.
    \label{eq:algentdefi}
\end{equation}
Note that $S_{i,m}^{\delta}\geq0$. The existence of this limit can be
proved in an analogous way as for finite dimensional maps, i.e.
using the sub-additivity of the logarithm and the so-called Fekete's lemma
\cite{Fekete1923}. When only a degree sequence is present, we will drop the
index $m$, and then write only $S_{i}^{\delta}$.

Then we give the following definition of integrability
based on the concept of algebraic entropy:

\begin{definition}
    A system of quad equations \eqref{eq:quadsys} is said to be
    \emph{directionally integrable} in the direction
    $\delta$ if $S_{i,m}^{\delta}=0$ for all $i=1,\dots,
    M$. If $S_{i,m}^{\delta}=0$ for all direction we say that the equation
    is \emph{integrable}.
    \label{def:intalg}
\end{definition}

\begin{remark}
    Before going on some remarks are in order:
    \begin{itemize}
        \item The rationale of defining (directonally) integrable those
            systems such that $S_{i,m}^{\delta}=0$ for all
            $i=1,\dots,M$ is that the system is expected to grow \emph{at
            the pace of the fastest one}. If the system can be decomposed
            in independent sub-systems this is clearly not a good measure
            of the growth of the sub-systems and they should be treated
            separately.
        \item The notion of directional integrability was elaborated in
            \cite{GKVHexEntropy}, even though in the cases that are
            presented in that paper it is not strictly needed (albeit
            theoretically possible). Some cases where this notion is actually
            \emph{needed} will be presented in \Cref{sec:examples}.  
        \item The condition $S_{i,m}=0$ implies that asymptotically as
            $l\to\infty$ the degree $d_{i,l,m}$ is \emph{sub-exponential},
            e.g.~ polynomial. Sometimes in the literature the linear
            asymptotic behaviour $d_{i,l,m} \sim l$ is considered to be
            related to \emph{linearisation}. This is usually true for quad
            equations, e.g. Boll's equations
            \cite{Boll2011,Boll2012a,Boll2012b}, see the discussion in the
            Introduction. Linear growth appeared also in
            \cite{GubKelsEntropy} for the type-$B$ face-centred quad
            equations satisfying the consistency around the face-centred
            cube property. However, at present, no explicit linearsation
            was produced for those systems. In \Cref{sec:examples} I will
            present some examples that show that for systems of quad
            equations the situation can be even more complicated.
    \end{itemize}
    \label{rem:seq}
\end{remark}

It is known, see \cite{BellonViallet1999,Viallet2015,Takenawa2001JPhyA},
that except for some pathological cases \cite{Hasselblatt2007} the sequence
of iterates of degree \emph{stabilises after a finite number of steps}. So,
in this paper I will aim to give a complete proof of the growth of the
examples I will be considering. I will resort to use an heuristic \emph{yet
extremely fast and efficient} method, which is to \emph{look for a
generating function for the sequences of degrees}. In particular, I will
make the \emph{ansatz} that the generating function is rational. This
allows me to find it algorithmically using Pad\'e approximants
\cite{BakerGraves1996}. The ability of fiding such a rational generating
function requires to compute a sufficiently high number of iterates, as
will be the case for all the example I will consider in
\Cref{sec:examples}.

To be more specific, given a sequence $\Set{d_k}_{k\in\N_0}$ its generating
function is the function admitting the sequence as Taylor coefficients:
\begin{equation}
    g(s) = \sum_{k=0}^{\infty} d_{k} s^{k}
    \label{eq:genfunc}
\end{equation}
In the case that such a generating function is rational one has two immediate
consequences:
\begin{itemize}
    \item The entropy is given by the inverse of the smallest modulus of
        the poles of $g(s)$, since the position of the smallest pole of $g$
        governs the asymptotics of its  Taylor coefficients, i.e. if
        $g(s)=P(s)/Q(s)$ then:
        \begin{equation}
            S  = 
        \log \left(\left[ \min\Set{ \vert s\vert\in\R^{+} \quad|\quad Q(s)=0}\right]^{-1}\right).
            \label{eq:algentgenfunc}
        \end{equation}
    \item The sequence $\Set{d_k}_{k\in\N_{0}}$ satisfies the constant
        coefficient recurrence relation:
        \begin{equation}
            Q(T^{-1}_{k}) d_{k} = 0, \quad T_{k}d_{k} = d_{k+1},
            \label{eq:Qeq}
        \end{equation}
        where we have written the polynomial $Q(s)$ as $Q(s) = 
        \sum_{j=0}^{J} a_{j}s^{j}$, $a_{0}=1$.
\end{itemize}
  
If all the poles of $g= P/Q$ lie on the unit circle, the denominator $Q(s)$
factorises as:
\begin{equation}
    Q\left( s \right) = \left( 1-s \right)^{\beta_{0}} 
    \prod_{i=1}^{K} \left( 1-s^{\beta_{i}} \right),
    \quad
    \beta_i \in \N,
    \label{eq:factor}
\end{equation}
the growth of the degree sequence is polynomial, and hence \emph{the
algebraic entropy vanishes}.

We resume some properties  of rational generating functions  in the
following results, obtained combining results from the books 
\cite[Chap. 6]{Elaydi2005}, \cite[Sect. 4.4]{Goldberg1986book}, and 
\cite{FlajoletOdlyzko1990}. For further discussion on the subject and more 
references we refer to the reviews  \cite{GubbiottiASIDE16, 
GrammaticosHalburdRamaniViallet2009}.

\begin{theorem}
    A sequence $\left\{ d_{k} \right\}_{k\in\N_{0}}$ admits a rational
    generating function $g\in\Cp\left( s \right)$ if and only if it 
    solves a \emph{linear difference equation with constant coefficients}.
    Moreover, if $\rho>0$ is the radius of convergence of $g$, writing
    $g$ as:
    \begin{equation}
        g = A\left( s \right) + B\left( s \right)\left( 1-\frac{s}{\rho} \right)^{-\beta},
        \quad \beta\in\N.
        \label{eq:gasy}
    \end{equation}
    where $A$ and $B$ are analytic functions for $\abs{s}<r$ such that
    $B(\rho)\neq 0$ we have:
    \begin{equation}
        d_{k} \sim \frac{B\left( \rho \right)}{\Gamma\left( \beta \right)}
        k^{\beta-1}\rho^{-k},
        \quad
        k \to \infty,
        \label{eq:dasy}
    \end{equation}
    where $\Gamma(s)$ is the Euler Gamma function.
    \label{th:rgf}
\end{theorem}

\begin{corollary}
    With the  hypotheses of \Cref{th:rgf} and the additional assumption
    that $\rho=1$, then
    \begin{equation}
        d_{k} \sim k^{\beta-1},
        \quad
        k \to \infty,
        \label{eq:dasyint}
    \end{equation}
    where $\beta=\beta_{0}+K$, with $\beta_0$ and $K$ as in
    \Cref{eq:factor} and $\gamma$ a constant factor.
    \label{cor:int}
\end{corollary}

In short, \Cref{th:rgf} and \Cref{cor:int} tell us that given a rational
generating function we can estimate the growth, which in the integrable case is
polynomial. In \Cref{sec:examples} we will make extensive use of these two
results.

\section{Examples}
\label{sec:examples}

In this section I will consider some examples of calculation of the
algebraic entropy for some known systems of quad equations. The result will
illustrate the various possibilities that were described in the previous
section. I decided to consider only two and three dimensional systems to
discuss these properties. As mentioned earlier I will denote the components
of such systems by $x_{l,m}$, $y_{l,m}$, and $z_{l,m}$.

\subsection{Coupled lpKdV systems}

Consider the following system of two quad equations:
\begin{subequations}
    \begin{gather}
        \left(x_{l,m}-x_{l+1,m+1}\right)\left(y_{l+1,m}-y_{l,m+1}  \right)
        -\alpha+\beta=0,
        \\
        \left( y_{l,m}-y_{l+1,m+1} \right)\left( x_{l+1,m}-x_{l,m+1} \right)
        -\alpha+\beta=0.
    \end{gather}
    \label{eq:cdpKdV}%
\end{subequations}
This system was presented in \cite{BradyXenitidis2022} where it was
discussed using the generalised symmetry method. It is a generalisation of
the lattice potential KdV equation, which is obtained from the reduction
$y_{l,m}=x_{l,m}$. Indeed, upon substitution the two equations collapse
into a single one:
\begin{equation}
    \left(x_{l,m}-x_{l+1,m+1}\right)\left(x_{l+1,m}-x_{l,m+1}  \right)
    -\alpha+\beta=0,
    \label{eq:dpKdV}
\end{equation}
which is known in the literature as the discrete potential KdV
equation\footnote{In the ABS classification \cite{ABS2003} this is the H1
equation.}. 

All directions are admissible for the system \eqref{eq:cdpKdV}, and the
system is strongly isotropic showing only the following degree sequence:
\begin{equation}
    1, 2, 4, 7, 11, 16, 22, 29, 37\dots
    \label{eq:degcdpKdV}
\end{equation}
The sequence \eqref{eq:degcdpKdV} has the following rational generating
function:
\begin{equation}
    g(s) =
    -\frac{s^2 - s + 1}{(s - 1)^3}.
    \label{eq:gfcdpKdV}
\end{equation}
From \Cref{cor:int} we see immediately that the growth is quadratic. In
this case we can be even more precise, and find the analytic form of the
sequence \eqref{eq:degcdpKdV} using for instance the
$\mathcal{Z}$-transform method \cite[Chap. 6]{Elaydi2005}:
\begin{equation}
    d_{k} = \frac{k(k + 1)}{2} + 1.
    \label{eq:dncdpKdV}
\end{equation}

Note that, the scalar equation \eqref{eq:dpKdV} is isotropic too, and in
particular has the same degree growth, see \cite{Tremblay2001}. For such an
equation the growth was proved rigorously using the $\gcd$-factorisation
method in \cite{RobertsTran2017}, see also \cite{vanderKamp2012}. So, the
system \eqref{eq:cdpKdV} behaves much like a scalar equation. This leaves
open the possibility of computing such growth rigorously using the
$\gcd$-factorisation method as was done in \cite{RobertsTran2017} for its
scalar version.

%[(x, [1, 2, 4, 7, 11, 16, 22, 29, 37]), 
%   (y, [1, 2, 4, 7, 11, 16, 22, 29, 37])]
%gf
%-(z**2 - z + 1)/(z - 1)**3
%coefficients dn
% n**2/2 + n/2 + 1

\subsection{Lattice NLS system}

As a second example we consider the following system:
\begin{subequations}
    \begin{gather}
        x_{l+1,m}-x_{l,m+1} - \frac{\alpha-\beta}{1+x_{l,m}y_{l+1,m+1}}
        x_{l,m} =0,
        \\
        y_{l+1,m}-y_{l,m+1} - \frac{\alpha-\beta}{1+x_{l,m}y_{l+1,m+1}}
        y_{l+1,m+1} =0,
    \end{gather}
    \label{eq:dNLS}%
\end{subequations}
This system was presented in \cite{KonstantinouRizos_etal2015} where it was
introduced as a discretisation of the nonlinear Schr\"odinger equation
(NLS), then considered again more recently in \cite{BradyXenitidis2022}.

The system \eqref{eq:dNLS} has $(+,+)$ and $(-,-)$ as admissible
directions. In those directions it is again strongly isotropic showing only
the degree sequence:
\begin{equation}
    1, 3, 7, 13, 21, 31, 43, 57, 73\dots
    \label{eq:degdNLS}
\end{equation}
The sequence \eqref{eq:degdNLS} has the following rational generating
function:
\begin{equation}
    g(s) =-\frac{s^2 + 1}{(s - 1)^3}.
    \label{eq:gfdNLS}
\end{equation}
From \Cref{cor:int} we see immediately that the growth is quadratic. In
this case we can be even more precise, and find the analytic form of the
sequence \eqref{eq:degdNLS} using for instance the $\mathcal{Z}$-transform
method:
\begin{equation}
    d_{k} = k(k + 1) + 1.
    \label{eq:dndNLS}
\end{equation}

%(('pp', [(x, [1, 3, 7, 13, 21, 31, 43, 57, 73]), 
%    (y, [1, 3, 7, 13, 21, 31, 43, 57, 73])]), 
%    ('mm', [(x, [1, 3, 7, 13, 21, 31, 43, 57, 73]), 
%    (y, [1, 3, 7, 13, 21, 31, 43, 57, 73])]))
%pp
%mm
%[('pp', [(x, -(s**2 + 1)/(s - 1)**3), (y, -(s**2 + 1)/(s - 1)**3)]), 
% ('mm', [(x, -(s**2 + 1)/(s - 1)**3), (y, -(s**2 + 1)/(s - 1)**3)])]
%[('pp', [(x, k**2 + k + 1), (y, k**2 + k + 1)]), 
% ('mm', [(x, k**2 + k + 1), (y, k**2 + k + 1)])]

\subsection{The lSG\textsubscript{2} and the lmKdV\textsubscript{2} systems}

In \cite{Hay2009} a completeness study of discrete systems on the
quad-graph arising from a $2\times 2$ Lax pairs was performed. Excluding
trivial, undetermined and overdetermined systems, the only two \emph{bona
fide} systems found in this classification were the lSG\textsubscript{2}
system:
\begin{subequations}
    \begin{gather}
        \begin{aligned}
        \frac{(\lambda^{(3)}_{l})^{(-1)^{m}}}{(\mu^{(3)}_{m})^{(-1)^{l}}} 
        \frac{x_{l,m+1}}{x_{l,m}}
        &+\lambda^{(1)}_{l}\mu_{m}^{(1)} x_{l+1,m+1}y_{l,m+1}
        =
        \\
        &\phantom{+}\frac{(\mu^{(3)}_{m})^{(-1)^{l}}}{(\lambda^{(3)}_{l})^{(-1)^{m}}} 
        \frac{x_{l+1,m+1}}{x_{l+1,m}}
        +\frac{\lambda^{(2)}_{l}\mu_{m}^{(2)}}{ x_{l,m}y_{l+1,m}},
        \end{aligned}
        \\
        \begin{aligned}
        \frac{(\mu^{(3)}_{m})^{(-1)^{l}}}{(\lambda^{(3)}_{l})^{(-1)^{m}}} 
        \frac{y_{l+1,m+1}}{y_{l,m+1}}
        &+\frac{\lambda^{(2)}_{l}\mu_{m}^{(2)}}{ x_{l,m+1}y_{l,m}}
        =
        \\
        &\phantom{+}\frac{(\lambda^{(3)}_{l})^{(-1)^{m}}}{(\mu^{(3)}_{m})^{(-1)^{l}}} 
        \frac{y_{l+1,m}}{y_{l,m}}
        +\lambda^{(1)}_{l}\mu_{m}^{(1)} x_{l+1,m}y_{l+1,m+1}
        \end{aligned}
    \end{gather}
    \label{eq:LSG2}%
\end{subequations}
and the lmKdV\textsubscript{2} system:
\begin{subequations}
    \begin{gather}
        \begin{aligned}
        \frac{\lambda^{(1)_{l}}}{(\mu^{(3)}_{m})^{(-1)^{l}}} 
        \frac{x_{l+1,m+1}}{x_{l,m+1}}
        &+\frac{\mu^{(2)}_{m}}{(\lambda^{(3)}_{l})^{(-1)^{m}}} 
        \frac{y_{l,m}}{y_{l,m+1}}
        =
        \\
        &\phantom{+}\lambda_{l}^{(2)} (\mu^{(3)}_{m})^{(-1)^{l}}
        \frac{y_{l,m}}{y_{l+1,m}}
        +(\lambda^{(3)}_{l})^{(-1)^{m}}\mu_{m}^{(1)}
        \frac{x_{l+1,m+1}}{x_{l+1,m}},
        \end{aligned}
        \\
        \begin{aligned}
        (\lambda_{l}^{(3)})^{(-1)^{m}}\mu_{m}^{(1)} 
        x_{l,m+1}y_{l+1,m+1}
        &+ \lambda_{l}^{(2)}(\mu_{m}^{(3)})^{(-1)^{l}} 
        x_{l,m}y_{l,m+1}
        =
        \\
        &\phantom{+}\frac{\mu_{m}^{(2)}}{(\lambda_{l}^{(3)})^{(-1)^{m}}}
        x_{l,m}y_{l+1,m} 
        \\
        &+
        \frac{\lambda_{l}^{(1)}}{(\mu_{m}^{(3)})^{(-1)^{l}}}
        x_{l+1,m}y_{l+1,m+1}.
        \end{aligned}
    \end{gather}
    \label{eq:LMKdV2}%
\end{subequations}
In \Cref{eq:LSG2,eq:LMKdV2} the functions $\lambda^{i}_{l}$ and
$\mu^{(i)}_{m}$ are \emph{arbitrary}. The name of the systems follows from
the fact that putting
\begin{equation}
    y_{l,m}=x_{l,m}, 
    \quad
    \lambda^{(i)}_{l}=\lambda^{(i)},
    \mu^{(i)}_{m}=\mu^{(i)},i=1,2,
    \quad
    \lambda^{(3)}_{l}=\mu^{(3)}_{l}=1,
    \label{eq:red}
\end{equation}
and applying a proper scaling of $x_{l,m}$, one gets the lattice
sine-Gordon (lsG)
equation \cite{QuispelCapel1991,Orfanidis1980}
\begin{equation} 
    x_{l,m}x_{l+1,m}x_{l,m+1}x_{l+1,m+1} =
    p(x_{l,m}x_{l+1,m+1}-x_{l+1,m}x_{l,m+1})+r
    \label{eq:lSG}
\end{equation}
and the lattice modified KdV (lmKdV) equation \cite{NijhoffCapel1995}
\begin{equation}
    x_{l+1,m+1} \left( px_{l,m+1}-r x_{l+1,m} \right)
    = x_{l,m} \left( p x_{l+1,m} - r x_{l,m+1} \right)
    \label{eq:lmKdV}
\end{equation}
respectively.

These systems are non-autonomous and contain arbitrary functions.  To
emulate such a behaviour in our computation of the degree sequences I
associate to any arbitrary function a random sequence of integers defined
on the range of the initial conditions. Running the computations more than
once I check that we obtain always the same result, meaning that it is
independent of the choice of the arbitrary functions.

Then, we have that the lSG\textsubscript{2} system has $(-,+)$ and $(+,-)$
as admissible directions. In such directions it is strongly isotropic with
the following sequence of degrees:
\begin{equation}
    1, 4, 10, 19, 31, 46, 64, 85, 109\dots.
    \label{eq:seqLSG2}
\end{equation}
This sequence of degrees has the following generating function:
\begin{equation} 
    g(s) =-\frac{s^2 + s + 1}{(s - 1)^3},
    \label{eq:gfLSG2}
\end{equation}
From \Cref{cor:int} we see immediately that the growth is quadratic. In
this case we can even more precise, and find the analytic form of the
sequence \eqref{eq:degcdpKdV} using for instance the
$\mathcal{Z}$-transform method:
\begin{equation}
    d_{k} = \frac{3}{2} k(k + 1) + 1.
    \label{eq:dnLSG2}
\end{equation}

Analogously, the lmKdV\textsubscript{2} system has $(-,+)$ and $(+,-)$ as
admissible directions, but differently from all other examples considered
up to now it is permutationally isotropic. For instance the degree
sequences in the $(-,+)$ direction are:
\begin{subequations}
    \begin{align}
        x_{l,m} &\colon 1, 4, 8, 15, 23, 34, 46, 61, 77\dots,
        \\
        y_{l,m} &\colon 1, 2, 6, 11, 19, 28, 40, 53, 69\dots,
    \end{align}
    \label{eq:seqLMKdV2}
\end{subequations}
while in the $(+,-)$ direction the two sequences are swapped. We have then
the following generating functions:
\begin{subequations}
    \begin{align}
        g_{x}(s) &= -\frac{s^3 + 2s + 1}{(s - 1)^3(s + 1)},
        \\
        g_{y}(s) &= -\frac{s^3 + 2s^2 + 1}{(s - 1)^3(s + 1)},
    \end{align}
    \label{eq:gfLMKdV2}%
\end{subequations}
which show, by \Cref{cor:int}, that the asymptotic growth is quadratic.
The factor $s+1$ in the denominators adds some two-periodic ``noise'' to
quadratic growth.

Let me now draw a brief comparison between the system and the scalar cases.
It is known \cite{Viallet2006,Tremblay2001} that the degree sequence of the
lSG equation \eqref{eq:lSG} is given by \eqref{eq:degdNLS} in all
directions, hence it is quadratic governed by \Cref{eq:dndNLS}. On the other
hand it is easy to see that the for all finite $k>0$ the degree of lSG
equation is
smaller than that of lSG\textsubscript{2} system:
\begin{equation}
    d_{k}(\text{lSG}) < d_{k}(\text{lSG\textsubscript{2}}).
    \label{eq:lSGvslSG2}
\end{equation}
In the same way it is known that the degree sequence of the lmKdV equation
\eqref{eq:lmKdV} is given by \eqref{eq:degcdpKdV} in all directions, hence
it is quadratic and governed by \Cref{eq:dncdpKdV}, so again the growth rates
at infinity are comparable. In this case it holds that
for all finite $k>1$:
\begin{equation}
    d_{k}(\text{lmKdV}) < d_{y,k}(\text{lmKdV\textsubscript{2}})<
    d_{x,k}(\text{lmKdV\textsubscript{2}}).
    \label{eq:vslmKdv2x}
\end{equation}

\subsection{Boussinesq-type systems}

In this final subsection, we consider three different kinds of
Boussinesq-type systems. Two of these systems are defined as ``incomplete''
systems, where one equation is a bona fide quad-equation and the other two
are defined only on three points. The latter one is a system of coupled
quad equations. A comprehensive review of these kind of systems is given in
\cite{HietarintaZhang2022}. We finally note that some results on the growth
of a Boussinesq equation defined on a $3\times 3$ stencil recently appeared
in \cite{Hietarinta2023Bous}.

\subsubsection{Boussinesq system}
The Boussinesq system is \cite{Nijhoff_etal1992}:
\begin{subequations}
    \begin{gather}
        z_{l+1,m} - x_{l,m}x_{l+1,m} + y_{l,m} =0,
        \label{eq:Bousqa}
        \\
        z_{l,m+1} -x_{l,m}x_{l,m+1} + y_{l,m} =0,
        \label{eq:Bousqb}
        \\
        (x_{l,m+1}-x_{l+1,m})(z_{l,m} -x_{l,m}x_{l+1,m+1} +y_{l+1,m+1})
        -p +q.
        \label{eq:Bousqc}
    \end{gather}
    \label{eq:Bousq}
\end{subequations}

Due to the fact that \Cref{eq:Bousqa,eq:Bousqb} are defined only on
the edges of the quad-graph \Cref{fig:quad}, we have that the only
admissible direction is the direction $(+,-)$. In such a direction the
degree sequences are the following:
\begin{subequations}
    \begin{align}
        x_{l,m} &\colon 1, 2, 4, 7, 14, 21, 30, 43, 55, 70, 89, 106, 127, 152, 
        174, 201, 232\dots,
        \\
        y_{l,m} &\colon 1, 2, 4, 9, 14, 21, 32, 43, 55, 72, 89, 106, 129, 
        152, 174, 203, 232\dots,
        \\
        z_{l,m} &\colon 1, 3, 5, 8, 15, 22, 32, 44, 56, 73, 90, 107, 131, 
        153, 175, 206, 233\dots.
    \end{align}
    \label{eq:seqBousq}
\end{subequations}
Then the generating functions have the following form:
\begin{equation}
    g_{i}(s) = -\frac{P_{i}(s)}{(s - 1)^3(s^2 + s + 1)^2}
    \label{eq:gfBousq}
\end{equation}
for $i=x,y,z$, where:
\begin{subequations}
    \begin{align}
        P_{x}(s) &= 4 s^6+3 s^5+5 s^4+s^3+2 s^2+s+1,
        \\
        P_{y}(s) &= 2 s^7+2 s^6+3 s^5+3 s^4+3 s^3+2 s^2+s+1,
        \\
        P_{z}(s) &= 5 s^6+3 s^5+3 s^4+s^3+2 s^2+2 s+1.
    \end{align}
    \label{eq:PxPyPzBous}%
\end{subequations}
Again from \Cref{cor:int} we have that all equations have quadratic growth.
The factor $(s^2 + s + 1)^{2}$ in the denominator of \Cref{eq:gfBousq}
introduces oscillations of order $k$.

\subsubsection{Schwarzian Boussinesq system}
The Schwarzian Boussinesq system is \cite{Nijhoff1996Dorfmann}:
\begin{subequations}
    \begin{gather}
        x_{l+1,m}y_{l,m} = z_{l+1,m} - z_{l,m},
        \label{eq:SchBousqa}
        \\
        x_{l,m+1}y_{l,m} = z_{l,m+1} - z_{l,m},
        \label{eq:SchBousqb}
        \\
        x_{l,m}y_{l+1,m+1}(y_{l+1,m}-y_{l,m+1}) 
        = y_{l,m} \left(p x_{l+1,m}y_{l,m+1}-q x_{l,m+1}y_{l+1,m}  \right).
        \label{eq:SchBousqc}
    \end{gather}
    \label{eq:SchBousq}
\end{subequations}

Due to the fact that \Cref{eq:SchBousqa,eq:SchBousqb} are defined only on
the edges of the quad-graph \Cref{fig:quad}, we have that the only
admissible direction is the direction $(+,-)$. In such a direction the
degree sequences are the following:
\begin{subequations}
    \begin{align}
        x_{l,m} &\colon 1, 4, 8, 12, 23, 34, 43, 62, 80, 94, 121, 146, 
        165, 200, 232, 256\dots,%, 299\dots,
        \\
        y_{l,m},z_{l,m} &\colon 1, 2, 7, 12, 19, 32, 43, 56, 77, 94, 113,
        142, 165, 190,227, 256\dots.%, 287\dots.
    \end{align}%
    \label{eq:seqSchBousq}
\end{subequations}
Then the generating functions have the same shape as \eqref{eq:gfBousq},
but with numerator polynomials indexed by $i=x,y|z$ given by:
\begin{subequations}
    \begin{align}
        P_{x}(s) &= 2 s^6 + 3 s^5 + 5 s^4 + 2 s^3 + 4 s^2 + 3 s + 1,
        \\
        P_{y|z}(s) &= 2 s^6 + 3 s^5 + 5 s^4 + 3 s^3 + 5 s^2 + s + 1.
    \end{align}
    \label{eq:PxPyz}%
\end{subequations}
Again from \Cref{cor:int} we have that all equations have quadratic growth,
with the same comment about the oscillations.

\subsubsection{Modified Boussinesq system}
The modified Boussinesq system is \cite{XenitidisNijhoff2012}:
\begin{subequations}
    \begin{gather}
        x_{l+1,m+1}(p y_{l+1,m}-q y_{l,m+1}) 
        = y_{l,m} (p x_{l,m+1}-q x_{l+1,m}),
        \label{eq:MBousqa}
        \\
        x_{l,m} y_{l+1,m+1} (p y_{l+1,m}-q y_{l,m+1}) 
        = y_{l,m} (p x_{l,m+1} y_{l,m+1}-q x_{l+1,m} y_{l+1,m}).
        \label{eq:MBousqb}
    \end{gather}
    \label{eq:MBousq}
\end{subequations}

All four directions are admissible for the system \eqref{eq:MBousq}.
Moreover, the system is permutationally isotropic. The sequence of the
degrees in the directions $(+,-)$ and $(-,+)$ the degree sequence is the
same, while in the directions $(+,+)$ and $(-,-)$ it is permuted. For
instance, in the $(+,-)$ direction we have the sequences:
\begin{subequations}
    \begin{align}
        x_{l,m} &\colon 1, 4, 8, 12, 23, 34, 43, 62, 80, 94, 121, 146, 
        165, 200, 232, 256\dots,%, 299\dots,
        \\
        y_{l,m} &\colon 1, 2, 7, 12, 19, 32, 43, 56, 77, 94, 113,
        142, 165, 190,227, 256\dots.%, 287\dots.
    \end{align}
    \label{eq:seqMBousq}
\end{subequations}
Then the generating functions have the same shape as \eqref{eq:gfBousq},
but with numerator polynomials indexed by $i=x,y$ given by:
\begin{subequations}
    \begin{align}
        P_{x}(s) &= (s^2 + 1) (s^4 + 2 s^3 + 3 s^2 + s + 1),
        \\
        P_{y}(s) &= s^6 + 2 s^5 + 4 s^4 + 2 s^3 + 4 s^2 + 2 s + 1.
    \end{align}
    \label{eq:PxPy}%
\end{subequations}
Again from \Cref{cor:int} we have that all equations have quadratic growth,
with the same comment about the oscillations.

%([(x, [1, 2, 6, 11, 17, 27, 37, 48, 64]), (y, [1, 3, 7, 11, 19, 29, 37, 51, 67])], 
%[(x, [1, 3, 7, 13, 21, 31, 43, 57, 73]), (y, [1, 3, 7, 13, 21, 31, 43, 57, 73])], 
%[(x, [1, 3, 7, 11, 19, 29, 37, 51, 67]), (y, [1, 2, 6, 11, 17, 27, 37, 48, 64])], 
%[(x, [1, 3, 7, 13, 21, 31, 43, 57, 73]), (y, [1, 3, 7, 13, 21, 31, 43, 57, 73])])    
%
%pm case (I'm just stupid I should have done the mp)
%"""

%\subsection{Boll's equations as systems}

\section{Discussion and perspectives}
\label{sec:concl}

In this short commemorative note I explained the construction I used over
the years to compute the algebraic entropy for systems of quad equations.
This construction follows closely the well known one for the scalar
setting, see \cite{Tremblay2001,Viallet2006}. My point of view is that the
system case needs to be handled more carefully because several properties
that are taken for granted in the scalar case no longer hold. Indeed, the main
problem is that for systems, multilinear it does not imply a well-defined
evolution in all directions any more.

I considered, the following examples:
\begin{itemize}
    \item a coupled lattice potential KdV system;
    \item a lattice NLS system;
    \item the lSG\textsubscript{2} and the lmKdV\textsubscript{2} systems;
    \item the Boussinesq system, and its Schwarzian and modified versions.
\end{itemize}
These systems showed several possible different behaviours. Some systems
behave much more like their scalar counterparts, like the coupled lattice
potential KdV system, while some others like the lSG\textsubscript{2} and
the lmKdV\textsubscript{2} systems do not. In particular, in the latter,
differently from the scalar case, the evolution is not defined in all
directions, and the lmKdV\textsubscript{2} is just permutationally
isotropic and not strongly isotropic. We also remark that the Boussinesq and
the Schwarzian Boussinesq systems have only one admissible direction.
However, using the ``augmentation'' procedure described in
\cite{Bridgman2013}, e.g. on the Schwarzian Boussinesq system
\eqref{eq:SchBousq} we obtain the following system:
\begin{subequations}
    \begin{gather}
        x_{l+1,m+1}y_{l,m+1} = z_{l+1,m+1} - z_{l,m+1},
        \label{eq:aSchBousqa}
        \\
        x_{l+1,m+1}y_{l+1,m} = z_{l+1,m+1} - z_{l+1,m},
        \label{eq:aSchBousqb}
        \\
        x_{l,m}y_{l+1,m+1}(y_{l+1,m}-y_{l,m+1}) 
        = y_{l,m} \left(p x_{l+1,m}y_{l,m+1}-q x_{l,m+1}y_{l+1,m}  \right).
        \label{eq:aSchBousqc}
    \end{gather}
    \label{eq:aSchBousq}
\end{subequations}
Without entering into the details, by a direct computation it is possible
to see that the system \eqref{eq:aSchBousq} has admissible directions
$(-,+)$, $(+,+)$, and $(-,-)$. In the direction $(-,+)$ the growth is still
quadratic, but the system becomes trivially linear in the other two
directions! So, the understanding of growth properties of systems of
quad equations and the definition of integrability seem to be more
complicated that in the scalar case.

Finally, we note that this method was originally conceived to check the
growth properties of trapezoidal $H_{4}$ equations and $H_{6}$ equations
\cite{Boll2011,Boll2012a,Boll2012b} written as systems of four
quad equations, see e.g.~\cite{GSL_general,GSY_DarbouxII}. These results
are too lengthy to be presented in this short note, but they can be
summarised by saying that all systems display linear growth,
and the $H_{4}$ ``systems'' have two admissible directions, while the
$H_{6}$ ``systems'' only have one. What are left out by this construction
are systems of quad equations where the explicit solution is algebraic in
\emph{all directions}. The set of integrable equations of this kind is
non-empty: in \cite{Kels2019Z} using a multi-component equivalent of the
Consistency Around the Cube procedure the following system, valid for all
$M\in\N$ was introduced:
\begin{equation}
    \sum_{i=1}^{M}
    \left(\frac{\alpha_{1}-\beta_{1}}{x_{l,m}^{(i)}-w_{k}}
    +\frac{\beta_{1}-\alpha_{2}}{x_{l+1,m}^{(i)}-w_{k}}
    +\frac{\alpha_{2}-\beta_{2}}{x_{l+1,m+1}^{(i)}-w_{k}}
    +\frac{\beta_{2}-\alpha_{1}}{x_{l,m+1}^{(i)}-w_{k}}\right)=0,
    \label{eq:kels}
\end{equation}
where $k=1,\dots,M$ and $w_{k}$, $\alpha_{j}$, and $\beta_{j}$ are
constants. Characterising the growth properties of such ``infinite
dimensional correspondences'' can be a challenging problem, which can
generalise the ideas given, for instance, in \cite{Truong2020,
Hone_etal2016}.

\subsection*{Acknowledgments}

I thank (alphabetical ordering) Giuseppe Gaeta, Orlando Ragnisco, Paolo Maria
Santini, and Federico Zullo for their help in handling the Special Issue of
OCNMP in Memory of Decio Levi, and their comments on the draft of this
manuscript.

Moreover, this work was supported by the GNFM INdAM through Progetto Giovani
GNFM 2023: ``Strutture variazionali e applicazioni delle equazioni alle
differenze ordinarie'' (CUP\_E53C22001930001), INFN Gruppo 4, and INFN Special
Initiative \emph{Mathemathical Methods of Nonlinear Physics}.

\printbibliography
%\begin{thebibliography}{99}
%
%
%\bibitem{Levi1}
%Levi D and Yamilov R I, The generalized symmetry method for discrete
%equations, {\it J. Phys. A: Math. Theor.} {\bf 42}, 454012 (18pp), 2009.
%
%
%\bibitem{Levi2}
%Levi D and Yamilov R I, Generalized Lie symmetries for Difference
%Equations. In: {\it Symmetries and Integrability of Difference Equations}. Ed.
%by D. Levi, P. Olver, Z. Thomova, and P. Winternitz. London Mathematical
%Society Lecture Notes series. Cambridge: Cambridge University Press, 160–190, 2011.
%
%
%\end{thebibliography}

\label{lastpage}
\end{document}